\documentclass[
10pt,
aps,
reprint,
nofootinbib,
showkeys,
longbibliography
]{revtex4-1}

\usepackage[utf8]{inputenc}
\usepackage{amsmath}
\usepackage{amsfonts}
\usepackage{amssymb}
\usepackage{graphicx}
\usepackage{graphics}
\usepackage{lipsum}
\usepackage{dcolumn}
\usepackage[dvipsnames]{xcolor,colortbl}
\definecolor{Gray}{gray}{0.85}
\usepackage{txfonts}
\usepackage[T1]{fontenc}
\usepackage[hidelinks]{hyperref}
\hypersetup{
	colorlinks   = true, 
	urlcolor     = blue, 
	linkcolor    = blue, 
	citecolor   = blue 
}

\begin{document}
	\renewcommand{\figureautorefname}{Fig.}
	\renewcommand{\equationautorefname}{Eq.}
	
	\title{Anisotropic strain effects in small-twist-angle graphene on graphite}
	\author{Márton \surname{Szendrő}} 
	\author{András \surname{Pálinkás}}
	\author{Péter \surname{Süle}} 
	\author{Zoltán \surname{Osváth}}
	\affiliation{Institute of Technical Physics and Materials Science (MFA), Centre for Energy Research, Hungarian Academy of Sciences, 1525 Budapest, P.O. Box 49, Hungary}
	
	\begin{abstract}
		
		The direct experimental probing of locally varying lattice parameters and anisotropic lattice deformations in atomic multilayers is extremely challenging. Here, we develop a new combined numerical/graphical method for the analysis of irregular moiré superstructures measured by scanning tunneling microscopy (STM) on a small-twist-angle ($\sim$0.6$^{\circ}$) graphene on highly oriented pyrolytic graphite (gr/HOPG). We observe distorted moiré patterns with a spatially varying period in annealed gr/HOPG. The nanoscale modulation of the moiré period observed by STM reflects a locally strained (and sheared) graphene with anisotropic variation of the lattice parameters. We use a specific algorithm based on a rigid lattice Fourier method, which is able to reconstruct the irregular and distorted moiré patterns emerging from strain induced lattice deformations. Our model is universal and can be used to study different moiré patterns occurring in two-dimensional van der Waals heterostructures. Additionally, room temperature scanning tunneling spectroscopy measurements show electronic states at the Dirac point, localized on moiré hills, which increase significantly the apparent corrugation of the moiré pattern. The measured topography is compared to classical molecular dynamics simulations. Density functional theory (DFT) calculations confirm that an AAB stacked trilayer region itself can contribute electronic states near the Fermi-level, in agreement with the measured peak in the local density of states. Furthermore CMD calculations reveal direction dependent bond alternations ($\sim$0.5\%) around the stacking regions, induced by shear strain, which could influence electronic properties.
		
	\end{abstract}

\date{\today}

\maketitle

\section{Introduction}
Graphene and other two-dimensional (2D) materials are intensively studied from the perspective of building new van der Waals heterostructures \cite{1} by stacking one layer on top of the other \cite{2,3,4,5,6}. The resulting new materials can reveal unusual properties and new phenomena \cite{7,8,9,10,11}. Therefore it is important to understand the interaction between stacked 2D layers and to study the resulting electronic properties. Already the interaction between two superimposed graphene layers implies rich physics and determines the properties of the stacked system \cite{12,13,14,15}. Moiré patterns can appear when two similar crystalline layers are superimposed, with a spatial period depending on the misfit and the rotation angle between the lattice parameters of the two layers \cite{16}. This moiré superstructure introduces not only a slight geometric corrugation in graphene, but also modulates the local density of states (DOS) \cite{12,13,14}. It was also shown that a moiré pattern can act as a perturbative periodic potential which induces secondary Dirac points in the graphene band structure \cite{17,18,19}. Graphene layers superimposed with small twist angle ($\sim$1$^{\circ}$) are of peculiar interest due to the decreased Fermi velocity and charge localization \cite{20}, the emergence of flat electronic bands \cite{21,22,23}, and also due to the role of electron-electron interaction \cite{23}. Recently, unconventional superconductivity and correlated insulator behavior was revealed for twist angles of around 1.1$^{\circ}$ \cite{24,25}. 
Furthermore, geometric and electronic effects induced by heterostrain in twisted graphene
layers near this magic angle were reported very recently \cite{26,34}. \par
\begin{figure*}[t!]
	\includegraphics[width=0.55\textwidth]{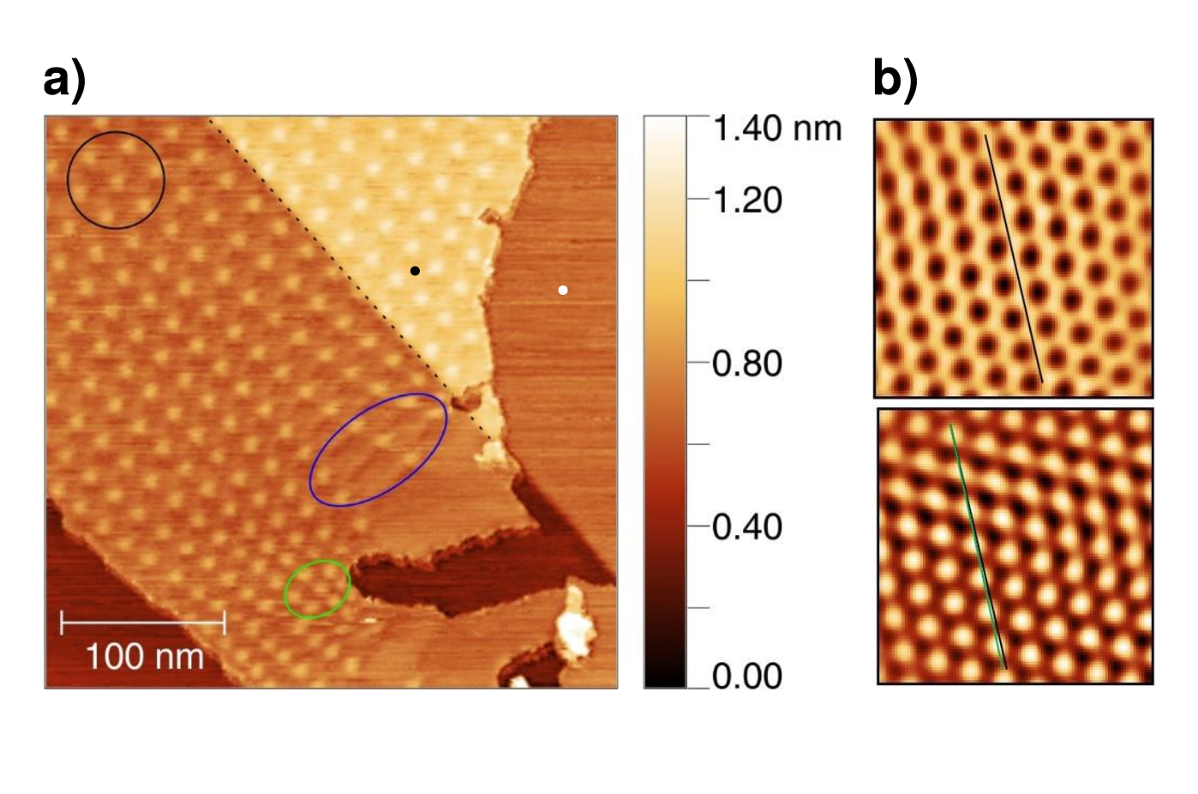}
	\caption{\label{fig:1}a) STM image of small-twist-angle graphene on HOPG. A moiré pattern with 22 nm period (black circle) is observed on graphene. The distance between moiré hills is position dependent: it decreases to 11-12 nm (green ellipse) and increases up to 52 nm (blue ellipse). Tunneling parameters: U = 100 mV, I = 0.25 nA. We marked the positions on graphene (black dot) and on HOPG (white dot) where the atomic resolution images shown in b) were measured. b) Atomic resolution images on graphene (top) and on HOPG (bottom). Graphene zig-zag orientation (black line) is nearly parallel to the HOPG zig-zag orientation (green line). Tunneling parameters: U = 100 mV, I = 1 nA.}
\end{figure*}

 
In this paper we investigate a small-twist-angle graphene on highly oriented pyrolytic graphite (HOPG) by scanning tunneling microscopy (STM) and show that the measured moiré superstructure reflects a locally strained graphene with anisotropic variations of the lattice parameter. Owing to the magnifying property of the moiré pattern, the system is perfectly suitable for studying strain effects and deformations. 
In theory, as the lattice mismatch and the twist angle between two adjacent layers tend to zero, the moiré wavelength tends to infinity, meaning that atomic scale deformations can be observed on the 10-100 nm scale moiré patterns. Indeed, this phenomenon is visible in our STM results, as the measured moiré becomes progressively irregular in one region of the sample (\autoref{fig:1}a). This is possible because in the measured system both magnifying criteria are fulfilled: the lattice mismatch is very small, and the twist angle is below 1$^{\circ}$. \par In addition, we show that the corrugation of the observed moiré pattern depends on the bias voltage used. Scanning tunneling spectroscopy (STS) measurements reveal a local DOS peak at the Dirac point, which is localized at the protruding sites of the moiré pattern.  This induces increased moiré corrugation when imaging at a bias voltage close to
the Dirac point (electronic effect). The results may have implications in the nanoscale engineering of the atomic and electronic properties of graphene and graphene based van der Waals heterostructures. \par
The paper is organized as follows. In Sec. II we summarize the experimental details of sample preparation as well as the applied density functional theory (DFT) methodology. The STM results and their interpretation is presented in Sec. III A, while in Sec. III B we provide a model to reconstruct the observed distorted patterns. In Sec. III C we discuss the charge localization observed at the moiré protrusions,  which is the main reason for the measured large apparent corrugation.  We compare the results with DFT calculations performed on AAB stacked trilayer graphene and a small-twist-angle moiré coincident cell. In Sec. III D we provide classical molecular dynamics (CMD) simulations for the observed small-twist-angle graphene on graphite system. The geometric corrugation obtained from simulations is compared to the experimental STM topography data. Finally, the conclusions are presented in Sec. IV. 

\section{Methods}

Graphene grown by chemical vapor deposition on copper foil \cite{26_b} was transferred onto HOPG substrate using thermal release tape (TRT). An etchant mixture consisting of CuCl2 aqueous solution (20\%) and hydrochloric acid (37\%) in 4:1 volume ratio was used to etch the copper foil. After etching, the TRT/graphene was rinsed in distilled water, then dried and pressed onto freshly cleaved HOPG. The TRT/graphene/HOPG stack was placed on a hot plate and heated to 100$^{\circ}$C, above the release temperature (90$^{\circ}$C) of the TRT. Thus, after one minute the TRT was easily removed, leaving behind the graphene on top of the HOPG. The sample was annealed at 650$^{\circ}$C in argon atmosphere for two hours, in order to remove residual contaminations from the TRT and to improve the adhesion of graphene to the HOPG. The graphene/HOPG system was investigated by STM and STS, using a DI Nanoscope E operating under ambient conditions. The STM measurements were performed in constant current mode. \par
\emph{DFT methodology:}  We used different levels of theroretical DFT methods for the simple trilayer graphene (TLG) calculations, and for the large scale calculation (commensurate moiré coincidence bilayer supercell including 33076 atoms).

a) For the AAB-TLG and ABA-TLG calculations (150 atoms/unit cell, aligned layers, zero twist) the accurate self-consistent DFT/GGA(PBE) method has been used used (Monkhorts-Pack 51x51x1, DZP numerical basis set, 300 Ry mesh cutoff, self-consistent field (SCF) tolerance is set to $10^{-6}$, SIESTA code \cite{48}). The periodic geometry of the ABA and AAB structures have been optimized by a conjugate gradient (cg) minimizer together with a supercell relaxation (variable cell method).

b) For the commensurate moiré coincidence bilayer (MCB) unit cell the non-self-consistent DFT/Harris functional (real space mesh cutoff 150 Ry \cite{48}) has been used with a single k-point ($\Gamma$-point calculation), which is sufficient for such a large system in which the Brillouin-zone and the reciprocal lattice become extremly small. This approach includes a single matrix diagonalization with fully calculated DFT matrix elements which provide the necessary eigenvalues for DOS calculation. The calculation of the projected DOS is extremely time-consuming for the MCB system, therefore we calculated only the total DOS, which is 
adequately informative on the features near the Fermi level. In both cases (a-b) the periodic superlattice structures have been pre-optimized (relaxed) using classical molecular statics (lammps code \cite{39}, see section III D). We used the lcbop/Kolgomorov-Crespi-z (lcbop/KC) hybrid potential \cite{40,41}  which includes the necessary van der Waals interaction term via the KC potential. The lcbop potential accurately describes graphitic systems in the planes and the inter-plane interaction has been simulated by the KC force-field (cohesive energy is in the range of a few tens of meV/C atom).

\section{Results}
\subsection{STM investigation of small-twist-angle graphene on HOPG }
An STM image of a graphene flake transferred onto HOPG is shown in \autoref{fig:1} a. The graphene passes over an atomic step of HOPG, which is marked with dashed line. A prominent feature of the STM image is the observed moiré pattern, the periodicity of which is around 22 nm (black circle) in most graphene areas. According to the formula 
$\lambda_{M} = a/[2\sin(\theta/2)]$ used for rotated graphene layers \cite{27,29}, where $a$= 0.246 nm is the graphene lattice constant, such a moiré period forms at a twist angle of $\theta$=0.64$^{\circ}$. Experimentally, we observe that the zig-zag orientations of graphene are nearly parallel with the zig-zag orientations of the HOPG substrate (\autoref{fig:1} b), which is in very good agreement with the above small twist angle. As one notices in \autoref{fig:1} a, the moiré periodicity is position dependent. There are areas where it decreases to 11-12 nm (green ellipse), and also where increases up to 52 nm (blue ellipse). This increase (decrease) is anisotropic, it occurs predominantly in one direction. Note that although the graphene edges have a shadow image due to a double STM tip, the moiré parameters are not affected (see Supplemental Material \cite{49}, Fig. S1).” 

\subsection{Reconstruction of the distorted moiré patterns using a rigid lattice Fourier method }
\begin{figure*}
	\includegraphics[width=0.71\textwidth]{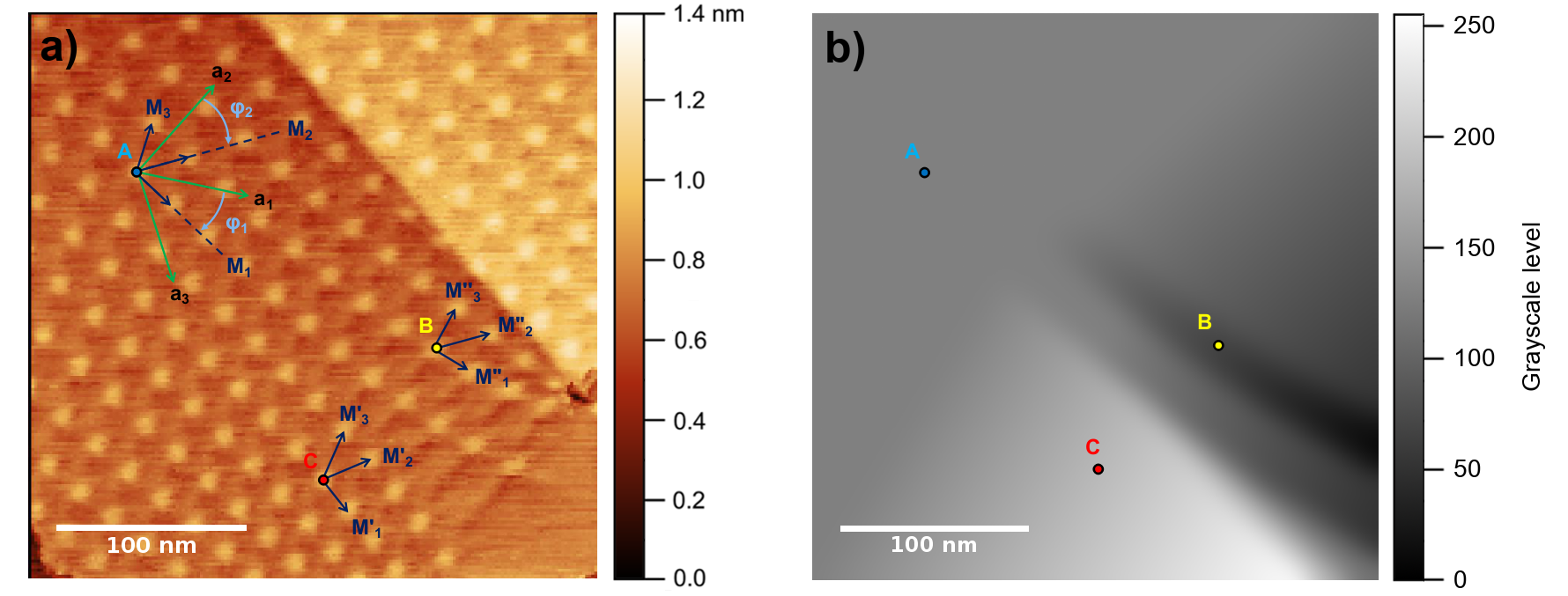}
	\caption{\label{fig:2} a) Experimental moiré pattern. The points $A, B, C$ denote areas where the moiré parameters were measured and used in the fitting procedure. $M_{1}, M_{2}, M_{3}$ are the local moiré wavelengths in the depicted directions. $a_{1}, a_{2}, a_{3}$ are the zig-zag directions of the graphene. $\varphi_{1}, \varphi_{2}$ are the moiré angles. b) The grayscale spatial map used in the RLFM model. Using linear interpolation, well defined graphene parameters were assigned to every grayscale level so that in $A, B, C$ the measured values were preserved.} 
\end{figure*}

In order to better understand the observed anisotropy, we developed a rigid lattice Fourier method (RLFM) to visualize the dependence of the moiré pattern on the spatial distribution of the graphene lattice constants. 
A model of a spatially varying anisotropic Moiré pattern can be found in \cite{26}. In their model the Authors used uniaxial strain with a varying twist angle as an approximation to heterostrain in a very specific setup. In our case, however, we have a random strain field, that arised from the transfer and annealing of the graphene layer. Therefore our model had to be very general. It can be used with any kind of lattices, with any type of strains, and can handle arbitrary lattice parameter gradients (see \autoref{fig:2} b) through our graphical method (see later). In the following we describe our model in detail.

A regular moiré pattern ($\xi$) can be defined by its Fourier-serie:
\begin{equation} \label{eq:1}
\xi(\vec{r}) = \sum_{nm} c_{nm} e^{i\left(n\vec{G}^{(M)}_{1}+m\vec{G}^{(M)}_{2}\right)\vec{r}}
\end{equation}
where $\vec{G}^{(M)}_{1}, \vec{G}^{(M)}_{2}$ are the reciprocal Moiré-vectors. We call the trigonal pattern \emph{isotropic} when the angle between the reciprocal moiré vectors is 120$^{\circ}$ and their length is the same. Deviations from these criteria make the pattern anisotropic. On a larger scale, $\vec{G}^{(M)}_{1}, \vec{G}^{(M)}_{2}$, can have spatial dependence $\vec{G}^{(M)}_{1} \equiv \vec{G}^{(M)}_{1}(x,y)$, $\vec{G}^{(M)}_{2} \equiv \vec{G}^{(M)}_{2}(x,y)$, in which case we call the pattern to be \emph{irregular}. The real space moiré vectors corresponding to $\vec{G}^{(M)}_{1}, \vec{G}^{(M)}_{2}$ are labeled $\vec{R}^{(M)}_{1}, \vec{R}^{(M)}_{2}$. According to Ref. \onlinecite{29}, the linear transformation between $\vec{R}^{(M)}_{1}, \vec{R}^{(M)}_{2}$ and the graphene lattice vectors $\vec{R}^{(O)}_{1}$, $\vec{R}^{(O)}_{2}$ is:

\begin{equation} \label{eq:2}
\left(\begin{array}{c}
\vec{R}^{(M)}_{1} \\ \vec{R}^{(M)}_{2}
\end{array} \right) = (1-M)^{-1}
\left(\begin{array}{c}
\vec{R}^{(O)}_{1} \\ \vec{R}^{(O)}_{2}
\end{array} \right)
\end{equation}

As \autoref{eq:2} shows, any anisotropy in the graphene lattice induces anisotropy also in the moiré pattern. In the work of Hermann \cite{29}, the theory is restricted to the case when anisotropy comes only from the length differences between
$\vec{R}^{(O)}_{1}$, $\vec{R}^{(O)}_{2}$. There, the angle between graphene lattice vectors $\omega_{O}$
and the angle $\omega_{S}$ between the substrate (HOPG) lattice vectors $\vec{R}^{(S)}_{1}, \vec{R}^{(S)}_{2}$ were considered to be the same ($\omega_{O} = \omega_{S} = \omega$), meaning that the author restricted the theory only to the cases where the same type of lattices are put above each other (e.g: hexagonal on hexagonal, rectangular on rectangular etc.). Here, however, we wanted to fully capture the anisotropic effects, therefore we allowed $\omega_{O}$ and $\omega_{S}$ to be different. Taking this into account we recalculated the matrix $M$, which was found to be:

\begin{widetext}
	\begin{equation} \label{eq:3}
	[1-M]^{-1} = \Gamma \left(\begin{array}{cc}
	\sin(\omega_{s})-p_{2}\sin(\omega_{o}+\alpha) & qp_{1}\sin(\alpha)  \\[1em]
	-\frac{p_{2}}{q}\sin(\Delta\omega + \alpha) & \sin(\omega_{s})-p_{1}\sin(\omega_{s}-\alpha) 
	\end{array}\right)
	\end{equation}
	
	where $\Gamma$ is:
	\begin{equation} \label{eq:4}
	\Gamma = \left[\sin(\omega_{s})-p_{1}\sin(\omega_{s}-\alpha)-p_{2}\sin(\omega_{o}+\alpha) + \frac{p_{1}p_{2}}{\sin(\omega_{s})}\left(\sin(\omega_{s}-\alpha)\sin(\omega_{o}+\alpha)+\sin(\Delta \omega + \alpha)\sin(\alpha)\right)\right]^{-1}
	\end{equation}
	
\end{widetext}

$p_{1} = R_{1}^{(O)}/R^{(S)}_{1}$, $p_{2} = R_{2}^{(O)}/R^{(S)}_{2}$, $q=R_{1}^{(S)}/R^{(S)}_{2}$ and $\Delta \omega = \omega_{O}-\omega_{S}$ and $\alpha$ is the twist angle between the graphene and the HOPG. We assume that there are no irregularities in the substrate, i.e. every deformation comes from the graphene (see Supplemental Material \cite{49}, Fig. S2). Thus, $\vec{R}^{(S)}_{1}, \vec{R}^{(S)}_{2}$ are the known equilibrium lattice vectors of graphite, and $\omega_{s} = 60^{\circ}$. Next, we consider the graphene lattice vectors as slowly varying, continuous functions of space: $\vec{R}^{(O)}_{1} \equiv \vec{R}^{(O)}_{1}(x,y)$, $\vec{R}^{(O)}_{2} \equiv \vec{R}^{(O)}_{2}(x,y)$. In this continuum model of graphene we have a well-defined lattice parameter in every point of the (x,y) plane.
If $\vec{R}^{(O)}_{1}(x,y)$ and $\vec{R}^{(O)}_{2}(x,y)$ are known, one can construct the moiré vectors $\vec{R}^{(M)}_{1}(x,y)$, $\vec{R}^{(M)}_{2}(x,y)$ using \autoref{eq:2}. Then, the moiré vectors are transformed to reciprocal space to get $\vec{G}^{(M)}_{1}(x,y)$, $\vec{G}^{(M)}_{2}(x,y)$ and the model moiré pattern $\xi(x,y)$ is obtained by building up the Fourier-serie of \autoref{eq:1}. The simplest moiré contains six Fourier-components with equal $c_{nm}$ coefficients. \par
The $[1-M]^{-1}$ matrix depends on the twist angle $\alpha$, which cannot be extracted directly from STM experiments. The measurable quantities which provide input for our model are the moiré wavelengths in different directions, and the moiré angle, $\varphi$, measured between the graphene zig-zag direction and the direction determined by neighboring moiré bumps. Furthermore, the moiré pattern is very sensitive on the $\omega_{O}$ parameter. It cannot be measured accurately enough from STM images, therefore we consider $\omega_{O}$ also an unknown parameter. To get around this problem of unknown parameters we implemented the least square fitting algorithm of Powell \cite{30}. The fitting parameters were $a_{1}$, $a_{2}$ (the length of the graphene lattice parameters), $\omega_{O}$, $\alpha$. The fitting data were the moiré wavelengths $M_{1}$, $M_{2}$, $M_{3}$, and two moiré angles $\varphi_{1},\varphi_{2}$ (\autoref{fig:2} a) (hereinafter \emph{moiré parameters}). These quantities can be calculated from the $\vec{R}^{(M)}_{1}, \vec{R}^{(M)}_{2}$ vectors for a particular set of $a_{1}$, $a_{2}$, $\omega_{O}$, $\alpha$ (hereinafter \emph{graphene parameters}), which can be fitted onto the measured data. \par 

\begin{figure*}[htbp]
	\includegraphics[width=0.85\textwidth]{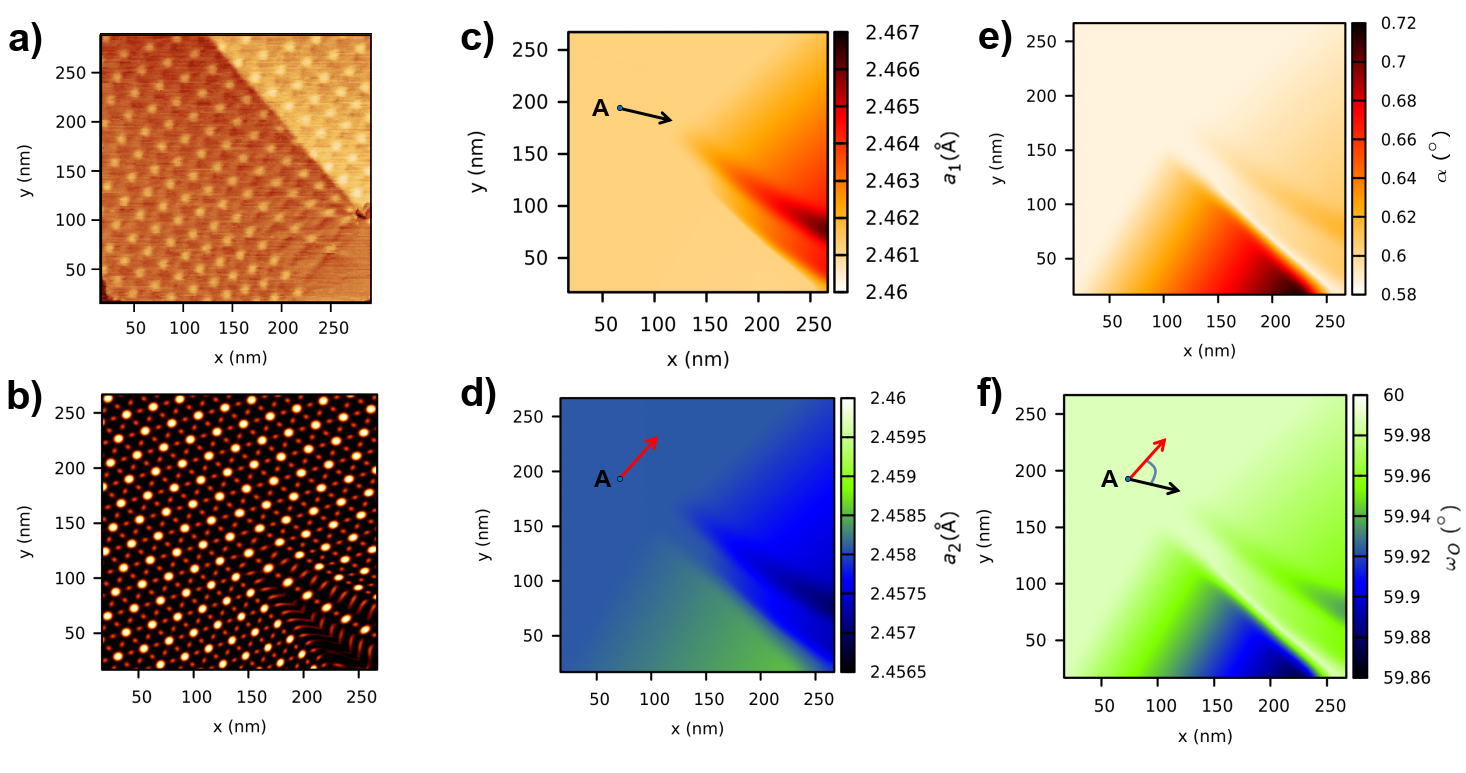}
	\caption{\label{fig:3} a) STM image of the moiré pattern. b) Simulated moiré pattern resulting from the RLFM model. c)-f) Spatial distributions of the following parameters: c) length of $a_{1}$ lattice parameter, d) length of $a_{2}$ lattice parameter, e) the twist angle $\alpha$ between the graphene and the HOPG, f) the angle between $a_{1}$ and $a_{2}$, $\omega_{O}$.The directions of $a_{1}$ and $a_{2}$ are marked in point A by black and red vectors, respectively.} 
\end{figure*}
\begingroup
\setlength{\tabcolsep}{10pt}
\renewcommand{\arraystretch}{1.5}
\begin{table*}
	\scriptsize
	\begin{tabular}{|c|c|c|c|c|c|c|}
		\hline 
		&  \multicolumn{2}{|c|}{A} &  \multicolumn{2}{|c|}{B} & \multicolumn{2}{|c|}{C} \\ 
		\hline 
		& STM & fit  & STM  & fit  & STM  & fit  \\ 
		\hline 
		$a_{1} (\text{\AA})$ & - & 2.461 & - & 2.467 & -  & 2.461 \\ 
		\hline 
		$a_{2} (\text{\AA})$ & - & 2.458 & - & 2.457 & - & 2.459 \\ 
		\hline 
		$a_{3} (\text{\AA})$ & - & 2.459 & - & 2.459 & - & 2.455 \\ 
		\hline 
		$\omega_{O} (^{\circ})$ & - & 59.99 & - & 59.93 & - & 59.87 \\ 
		\hline 
		$\alpha (^{\circ})$ & - & 0.59 & - & 0.62 & - & 0.72 \\ 
		\hline
		$M_{1} (\text{\AA})$ & 231 & 230.92 & 220 & 218.42 & 228 & 229.89 \\ 
		\hline 
		$M_{2} (\text{\AA})$ & 225 & 225.03 & 195 & 194.87 & 187 & 185.66 \\ 
		\hline 
		$M_{3} (\text{\AA})$ & 252 & 252.04 & 270 & 271.34 & 224 & 222.32 \\ 
		\hline 
		$\varphi_{1} (^{\circ})$ & 28 & 28.29 & 20 & 18.71 & 35 & 35.46 \\ 
		\hline 
		$\varphi_{2} (^{\circ})$ & 31 & 30.71 & 32 & 25.81 & 21 & 27.41 \\ 
		\hline 
	\end{tabular} 
	\caption{\label{tbl:1} Result of the fitting procedure. For each point ($A,B,C$) the table contains the fitted graphene-parameters ($a_{1},a_{2},a_{3},\omega_{O},\alpha$) and the corresponding Moiré-parameters ($M_{1},M_{2},M_{3},\varphi_{1}, \varphi_{2}$) comparing it with the measured Moiré-parameters ('STM' column).}
\end{table*}
\endgroup 
\begin{figure*}
	\includegraphics[width=0.83\textwidth]{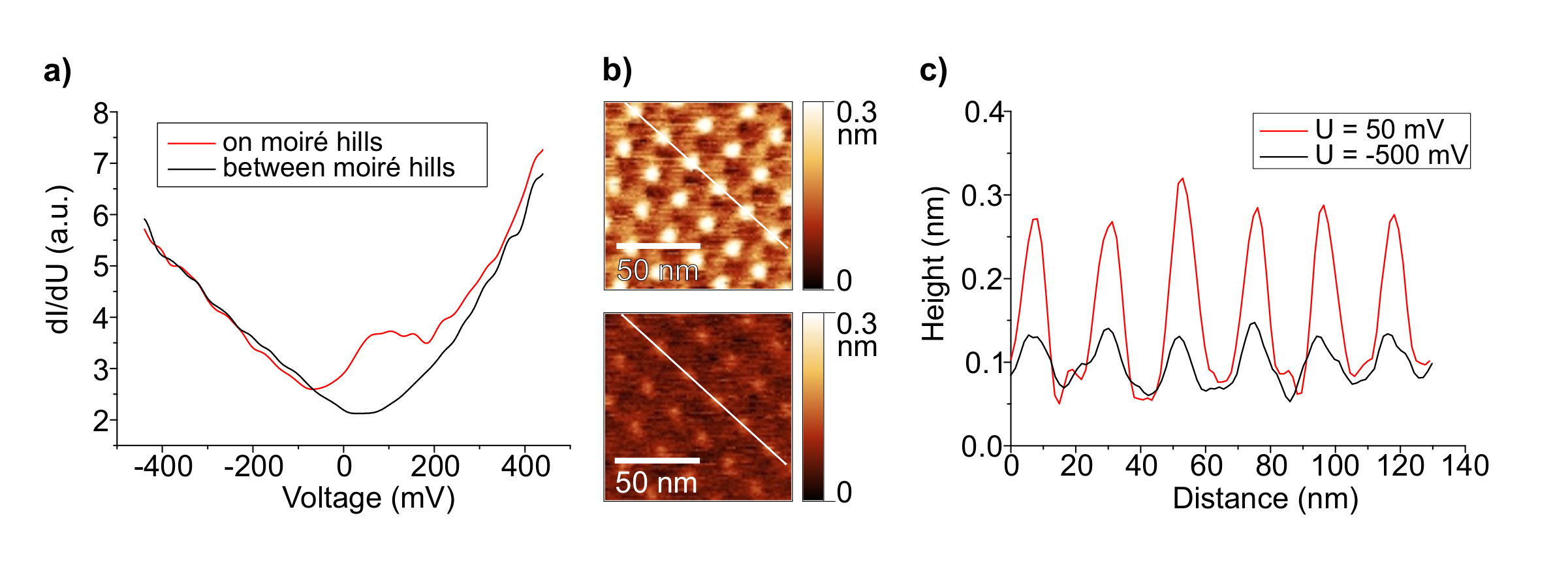}
	\caption{\label{fig:4}(a) dI/dU spectra measured on moiré hills (red curve) and moiré valleys (black curve). (b) STM images of the same moiré pattern measured with bias voltages of U = 50 mV (top panel) and U = -500 mV (bottom panel). (c) Height profiles taken at U = 50 mV (red) and U = -500 mV (black) along the same moiré hills, marked with white lines in b).}
\end{figure*}
In particular, we measured the moiré parameters in three different locations marked with 
$A,B,C$ in the STM topography image of \autoref{fig:2} a. The three locations were chosen to capture the basic irregularity of the moiré pattern: the bending of the moiré rows in opposite directions. In $A$ the pattern is slightly anisotropic, but it is regular. As one goes along a straight path from $A$ to $B$, the pattern is rotating with a positive angle, while it is rotating with negative angle through a path from $A$ to $C$. This makes the pattern highly irregular in the region between $B$ and $C$. \par
Once the moiré parameters are known, the graphene parameters are fitted in each location. Then, the spatial dependence of these graphene parameters $a_{1}(x,y),a_{2}(x,y),\omega_{O}(x,y),\alpha(x,y)$ has to be built. If these are known, $\vec{R}^{(O)}_{1}(x,y)$ and $\vec{R}^{(O)}_{2}(x,y)$ and $[1-M](x,y)^{-1}$ can be formulated, and the anisotropic, irregular moiré pattern model can be visualized. Our approach here was the following: a grayscale spatial map was generated using gradient tools in a graphic editor software (\autoref{fig:2} b), and in a separate step, linear interpolation was used in our developed code to assign every grayscale level of the pixels a graphene parameter value, such that the fitted values in the $A,B,C$ points were preserved. In this way we were able to conjecture the spatial distribution of the graphene parameters, i.e. the distortion in the graphene lattice. As an approximation, we used the same grayscale map for each graphene parameter (with different parameter-grayscale-level interpolation), meaning that we assumed the same basic spatial distribution for the lattice parameters and for the twist angle. The moiré pattern was then constructed from $a_{1}(x,y),a_{2}(x,y),\omega_{O}(x,y),\alpha(x,y)$ using \autoref{eq:1}. The next step was to manually fine-tune the grayscale map in a trial and error manner by comparing the simulated moiré pattern with the STM image until a high level of agreement was attained. The simulated moiré pattern is shown in \autoref{fig:3} b. As a result of our model the deformation of the graphene is revealed in details in \autoref{fig:3} c-f.

It can be seen that the deformations of the lattice parameters are highly anisotropic (\autoref{fig:3} c-d). For example, between points $A$ and $B$, there is a contraction of 0.21\% in $a_{2}$, but a dilatation of 0.27\% in $a_{1}$. We were able to determine the spatial dependence of the twist angle $\alpha$ (\autoref{fig:3} e), which is not a trivial task for irregular and anisotropic patterns. For an isotropic and regular moiré pattern with wavelength of 22.2 nm the twist angle is 0.64$^{\circ}$ in equilibrium. We found slightly different values in points $A$, $B$, and $C$, namely 0.58$^{\circ}$, 0.62$^{\circ}$, and 0.72$^{\circ}$, respectively. The deviations from the equilibrium value are clearly very small, however they represent a crucial factor in the deformation of the moiré pattern. Because of the strong magnifying property in this special system, very small changes in the graphene parameters induce prominent effects on the scale of tens of nanometers. To illustrate this, we recalculated the whole moiré pattern by changing $\alpha$ in the points $B$ and $C$ by a small amount, but keeping the other graphene parameters at their fitted value. A change of only 0.1$^{\circ}$ in the twist angle resulted in a deflection of the moiré rows by 8$^{\circ}$, which is a significant effect. A similar sensitivity is true for $\omega_{O}$ as well. We found that $\omega_{O}$ is slightly less than 60$^{\circ}$: 59.98$^{\circ}$, 59.93$^{\circ}$, and and 59.86$^{\circ}$ in the points $A$, $B$, and $C$, respectively. If $\omega_{O}$ was 60$^{\circ}$, as it would be in graphene without shear deformation, the deflection of moiré rows would change by 7$^{\circ}$. This clearly shows the sensitivity of the moiré pattern on the graphene parameters. The calculated graphene parameters for the points $A$, $B$, and $C$ are summarized in \autoref{tbl:1}, were additionally, the calculated moiré parameters are compared with the measured ones. The observed deformations and deviations from equilibrium values are ascribed to local strain and twist inhomogenity in graphene, probably induced during annealing.

\begin{figure*}
	\includegraphics[width=0.60\textwidth]{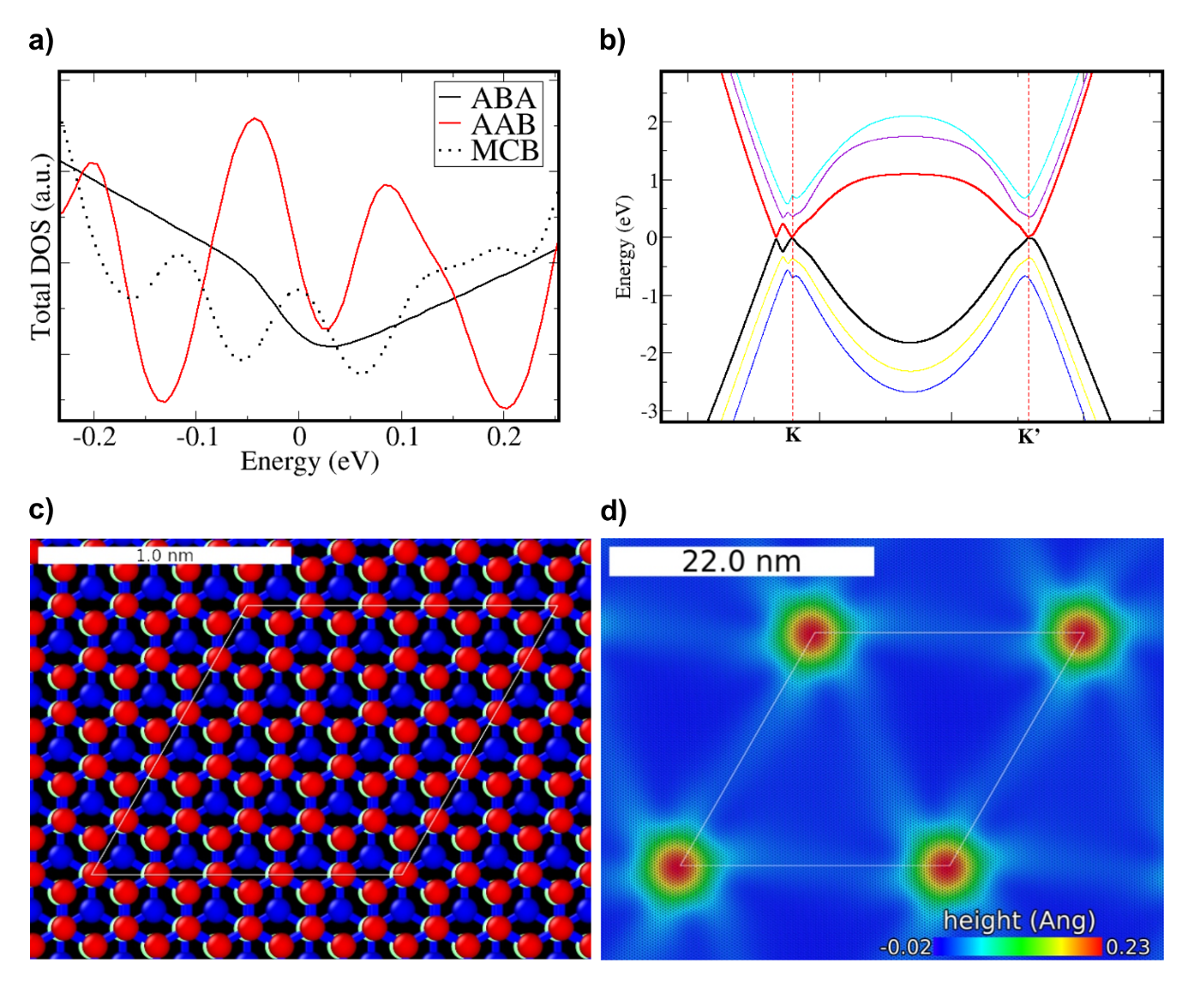}
	\caption{\label{fig:5} DFT calculations of small-angle twisted TLG with different stackings, compared with the moiré commensurate bilayer (MCB) supercell (33076 atoms). (a) Calculated total DOS for ABA-TLG (black), AAB-TLG (red), and for the MCB supercell at misorientation angle of 0.63$^{\circ}$.  (b) Calculated band structure for the AAB-TLG. (c) The relaxed superlattice of the AAB-TLG system with 150 atoms. (d) The topography plot of the relaxed superlattice of the MCB. In both cases (c-d) the periodic superlattices (AAB-TLG, ABA-TLG, MCB) have been optimized by lcbop/Kolgomorov-Crespi potentials \cite{40,41}. Gaussian broadening of 0.04 eV was used in all DOS calculations.\textbf{}}
\end{figure*}

\subsection{Charge localization on moiré hills}
In the following, we will study the local DOS of graphene. STS measurements were performed far from edges, on a graphene part with regular moiré pattern (\autoref{fig:4} a). The dI/dU spectra show a slightly p-doped graphene with the Dirac point at $U_{D}$$\approx$50 mV, as observed from the spectrum measured in a moiré valley (black line). 
In contrast with this typical graphene-like spectrum, the dI/dU measured on moiré hills reveal significantly higher local DOS near the Dirac point. Such peak localization can be induced by the moiré potential \cite{31,32}, and was observed in low temperature STM/STS measurements of twisted graphene layers with small angle \cite{33,34}. We note that at such low twist angle the moiré induced van Hove singularities are very close to each other \cite{33} and they are not resolved anymore, as the Fermi velocity tends to zero \cite{31}. \par 

In order to understand in more detail the observed local DOS peak, DFT calculations have been carried out for CMD pre-optimized  TLG supercells with AAB and ABA stackings, which take into account also the uppermost two layers of the HOPG substrate. Full self-consistency calculations performed on AAB-TLG supercell resulted in a band structure as shown in Fig. 5b. The corresponding DOS calculations (\autoref{fig:5} a, red) show a double-peak feature near the Dirac point. Such two-peak DOS structure is expected to occur at twist angles below 1$^{\circ}$ from continuum model calculations as well \cite{32}. Here, a Gaussian broadening of 0.04 eV was used to account for the room temperature used in the STS experiments, which actually smears the double-peak feature.  Note that the DOS of ABA-TLG is featureless near the Fermi energy (\autoref{fig:5} a, black). These results capture very well the peak localization in the AA stacked regions and are in good agreement with the measurements, although the peak splitting is not well resolved by room temperature STS (Fig. 4a). 

In addition, we performed also a constrained, Harris-functional-like DFT calculation for the full CMD pre-optimized  commensurate MCB supercell, including 33076 atoms at 0.63$^{\circ}$ twist angle (\autoref{fig:5} d).  In this case a single diagonalization of the Hamiltonian provides us the eigenvalues for total DOS calculation. This approach goes beyond the level of simple (even DFT fitted) tight-binding methods since the DFT matrix elements are calculated at first-principles level. The obtained DOS (\autoref{fig:5} a, dashed line) is comparable with the more accurate DFT calculations of the AAB-TLG supercell. We find that the MCB supercell gives a sharp single peak at the Fermi level, in agreement with the flat band reported for small-angle twisted graphene bilayers \cite{32,36}. No peak splitting occurs for the MCB commensurate supercell at 0.63$^{\circ}$ twist angle, which could be related to the mixed effect of differently stacked regions. The detailed analysis of the MCB DOS spectral features goes, however, beyond the scope of the present article. 
STM simulation of the MCB supercell is challenging as the non-self-consistent Harris charge density yields poor surface energies \cite{50}, and the available basis set for such calculation is short ranged \cite{48}.  Thus, the simulated STM images lack of sufficient contrast to display the moiré pattern and the associated small charge density modulations. 
	

\par 
The apparent corrugation of the moiré pattern depends strongly on the bias voltage (U) used in the STM measurements. This effect is clearly observed in \autoref{fig:4} b, where the moiré hills appear significantly much brighter at U=50 mV (top panel) than at U=-500 mV (bottom panel). Note that the vertical scale is the same in the two images. Line profiles taken through the same bumps (marked with white lines) reveal that the moiré hills are 2-3 times higher for U=50 mV (\autoref{fig:4} c). This dependence of the apparent height on the bias voltage is closely related to the localized DOS peak, which contributes significantly to the tunneling current. Hence, at U=50 mV
the STM tip has larger upwards (z) movements in order to keep the current constant when scanning the moiré hills. In turn, at U=-500 mV the local DOS of moiré hills and valleys is similar, therefore the electronic contribution to the measured corrugation is less significant, and the z-movement of the STM tip can be attributed mainly to the geometric corrugation.

\begin{figure*}
	\includegraphics[width=0.9\textwidth]{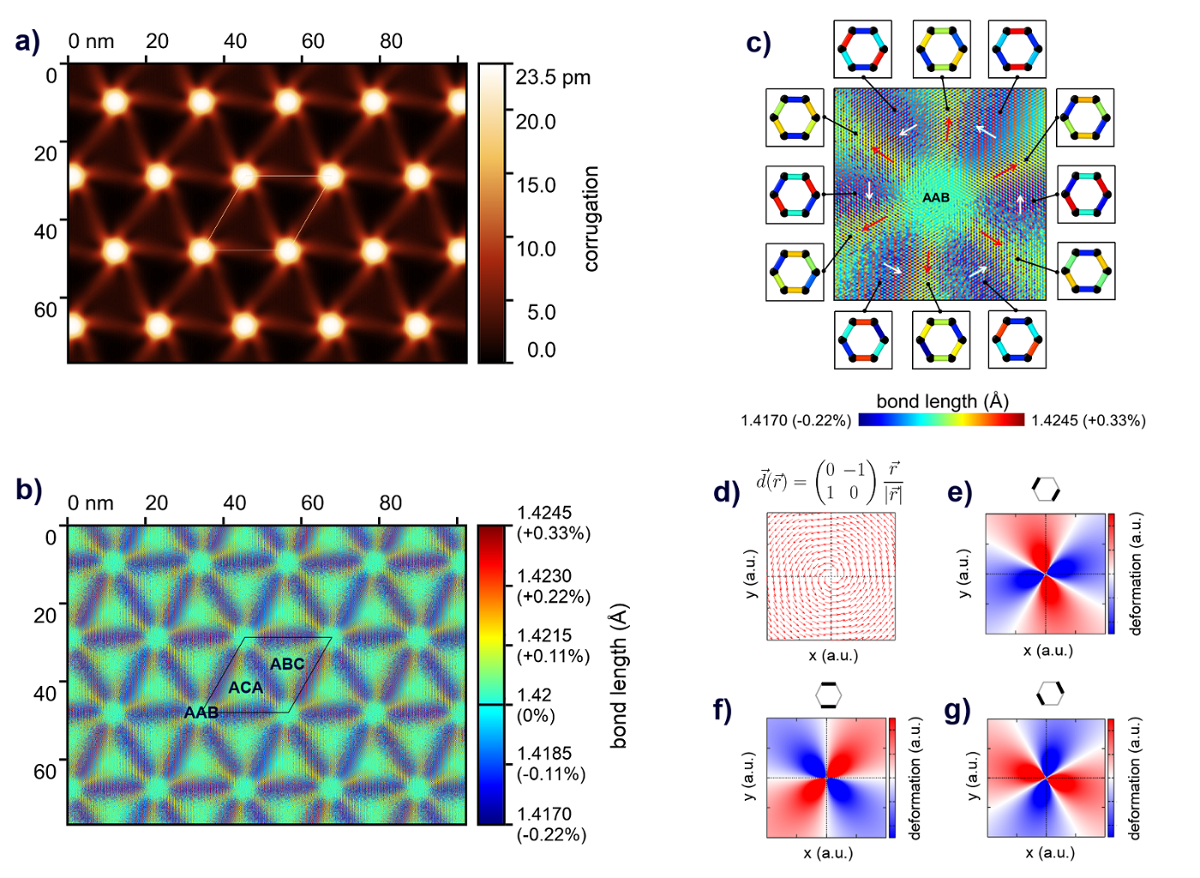}
	\caption{\label{fig:6} a) Topography of graphene for a 22.37 nm $\times$ 22.37 nm commensurate rhombic moiré superlattice at 0.63$^{\circ}$ twist angle, for graphene on 22 monolayer thick HOPG, as computed by classical molecular dynamics geometry relaxation. (b) Color coded bond length map of the Moiré superlattice. In all cases the rhombic simulation cells are denoted with solid lines. c) Fine structure of bond alternations around an $AAB$ stacking from the simulation depicted in a)-b). The colored hexagons show how the graphene lattice is deformed in average in that particular region. Red and white arrows indicate the main zig-zag directions, where the bond alternations are the most pronounced. d) Simple analytic homogeneous twisting displacement field $\vec{d}(\vec{r})$. e-g) Direction dependent length change of the bonds marked with the black thick lines in the hexagons above the figures induced by the displacement field in d).} 
\end{figure*}

\subsection{CMD simulations of small-twist-angle graphene on HOPG}
In order to support these findings, we carried out classical molecular dynamics simulations of a small-twist-angle graphene/HOPG system with regular and isotropic moiré pattern. For this purpose, we used a commensurate supercell, which was selected in a manner to approach the experimental moiré parameters and twist angle. This supercell is a $\sqrt{8269}$ graphene $\times$ $\sqrt{8269}$ HOPG structure, which was obtained by iterating through highly ordered commensurate cells as described in Refs \onlinecite{37,38}. The cell has a twist angle of 0.63$^{\circ}$, a moiré wavelength of 22.37 nm, and a moiré angle of 29.68$^{\circ}$. The CMD simulation was implemented in the LAMMPS code \cite{39}. \par
CMD simulations for bilayer graphene can be found in the literature \cite{28,42,47}, and also for graphene on graphite \cite{28}. In the latter case  only one rigid layer of graphene was used to mimic the substrate. However, at least a third layer should be considered as it is necessary to reproduce the proper stackings ($AAB$, $ACA$, $ABC$) occurring in the Moiré-pattern of a gr/HOPG system. In STM measurements the effect of the third layer is not negligible \cite{46} (although it is mainly an electronic effect).
Therefore we placed the graphene above a thick HOPG substrate, which consists of 22 carbon layers.  The long range bond order potential was utilized \cite{40} for carbon-carbon interactions within a single carbon layer.
The weak van der Walls forces between the carbon layers were modeled using the Kolmogorov-Crespi potential \cite{41}. 
We performed geometry relaxation using the FIRE algorithm. \par
The resulting topography of the graphene sheet is shown in \autoref{fig:6} a.
The maximal geometric corrugation induced by the moiré pattern obtained from CMD is 0.23 Å. This is even smaller than the corrugation of the experimentally found moiré pattern measured at U=-500 mV (0.6 – 0.8 Å, \autoref{fig:4} c). The combination of the often-used carbon potentials (lcbop, rebo, airebo) with the Kolmogorov-Crespi potential might underestimate somewhat the corrugation \cite{28}. However, the correction stemming from this underestimation does not compensate for the measured large apparent height. In light of this, the large apparent height of the moiré hills measured at 
U=50 mV can only be understood by taking into account significant electronic effects, as described above. \par  

The carbon-carbon bond length distribution in the optimized moiré supercell (see \autoref{fig:6} b) shows that the local strain
is between -0.22\% and +0.33\%. The color coded pattern reveals some fine details of the moiré superlattice. 
The pattern consists of three green spots and two types of interconnections
between them: a purple and a yellow one (\autoref{fig:6} b). The green
spots are the different stacking regions ($AAB$, $ACA$, $ABC$) having nearly
equilibrium bond length ($\sim$1.42 Å), inferring a period of $\sqrt{3}/3\lambda_{M}$ in the bond length distribution. Due to the
local twisting of the top graphene layer around the stacking
regions \cite{28,42,43,47}, there is a spatially changing local shear around them, which manifests in rich bond alternations (\autoref{fig:6} c). 
The most pronounced bond alternations occur in the purple areas which interconnect the $AAB$ regions.
These regions appear purple because they are mixed up with red (dilated) and blue (contracted) bonds (\autoref{fig:6} b-c). It can be easily seen that these regions coincide with the so-called soliton walls visible on the CMD simulated topography (\autoref{fig:6} a), previously reported in twisted bilayer graphene \cite{44}. The yellow areas connect
different stackings: the $AAB$ with $ACA/ABC$.
\par
In \autoref{fig:6} c) we analyzed the fine details of the bond alternations around the $AAB$ stacking. The nine colored hexagons show how the graphene lattice is deformed in average in that particular region. In each interconnection a major zig-zag direction can always be identified, where the bond alternation effect is the strongest. We indicated these major directions with white and red arrows in \autoref{fig:6} c. 
In the case of purple interconnections it turned out that the main direction for the bond alternation is the zig-zag direction perpendicular to the soliton wall (white arrows). In contrast, for the yellow interconnections the major direction is parallel with the interconnection itself (red arrows).
The twisting displacement field around the stacking has the property that it deforms the least those bonds that are either parallel (light green bonds in yellow interconnections) or perpendicular (light blue bonds in purple interconnections) to the local displacement field. We depicted this in \autoref{fig:6} d-g), where we showed how the bonds in specific directions would deform in a very simple analytic homogeneous twisting diplacement field $\vec{d}(\vec{r})$ (see \autoref{fig:6} d)). Although this is an oversimplification as the real field has a radial dependence and the displacements are not necessarily perpendicular to the $(x,y)$ position vectors of the carbon atoms \cite{47}, it gives a very good insight what actually happens with the bonds. The two-fold symmetry observed in \autoref{fig:6} e-g)  is modulated on a larger scale by the six-fold symmetry of the Moiré-pattern (\autoref{fig:6} b-c).  
The bond pattern in \autoref{fig:6} c) has a point reflection symmetry around the AAB stacking. 
Such a specific lattice arrangement could influence the electronic properties through pesudo-magnetic fields as the hoppings are having a very unique directional feature. If this twisting deformation could be enhanced with external techniques up to the critical shear strain ($\sim$ 17\%) for gap opening, it could lead to direction dependent gaps, meaning that in a well set configuration the graphene would conduct electrons only between certain stackings, that could lead to localization effects and flat-band physics independently of the magic-angle.

\section{Conclusion}
Small-twist-angle graphene on HOPG has been studied both experimentally and theoretically. The moiré superlattices observed by STM reflected a locally strained graphene with anisotropic variations of the lattice parameter. We developed a combined graphical-numerical method in order to evaluate the deformations that resulted in these distorted moiré patterns. As a result of our new approach the spatial dependence of the anisotropic deformations was revealed in unprecedented detail: not only the anisotropic moiré pattern could be reproduced, but also the local values of ($a_{1}, a_{2}, \omega_{O}, \alpha$) could be accurately calculated. The sensitivity of the moiré pattern on the variation of graphene parameters was also demonstrated. Additionally, a local DOS peak at the Dirac point was observed, localized at the protruding sites of the moiré pattern, which resulted in a significant increase of the apparent moiré corrugation. These findings were supported by classical molecular dynamics simulations, which also revealed rich bond alternation patterns around the stackings, induced by shear strain, which could have interesting applications in the future of strain engineering.  DFT calculations confirmed that the measured local DOS peak can be attributed to $AAB$ stacked trilayer regions in small-twist-angle gr/HOPG systems. The results may have implications in the nanoscale strain engineering of the atomic and electronic properties of graphene based van der Waals heterostructures.

\section*{ACKNOWLEDGMENTS}
We acknowledge financial support from the National Research, Development and Innovation Office (NKFIH) through the OTKA Grants K-119532 and KH-129587, as well as from the Korea-Hungary Joint Laboratory for Nanosciences.


%

\end{document}